\documentclass[aps,prb,superscriptaddress,preprintnumbers,
showpacs,a4,twoside,twocolumn]{revtex4}

\usepackage{bm}
\usepackage{graphicx}
\usepackage{amsfonts}
\usepackage{amsmath}
\usepackage{amssymb}
\usepackage{times}
\usepackage{color}  

\begin{document}

\title{Comment to ``Critical Exponents of the Superfluid-Bose Glass Transition in Three-Dimensions" by Z. Yao et al., arXiv:1402.5417v1}

\author{Rong Yu}
\affiliation{Department of Physics, Renmin University of China, Beijing 100872, China}
\author{Vivien S. Zapf}
\affiliation{National High Magnetic Field Laboratory, Los Alamos National Laboratory, Los Alamos, New Mexico 87545, USA}
\author{Tommaso Roscilde}
\affiliation{Laboratoire de Physique, CNRS UMR 5672, Ecole Normale Sup\'erieure de Lyon, Universit\'e de Lyon, 46 All\'ee d'Italie, 
Lyon, F-69364, France, and Institut Universitaire de France}

\begin{abstract} In this comment we address the preprint Z. Yao et al. (arXiv:1402.5417v1) concerning the conditions upon which the correct critical { temperature} exponent $\phi$ can be observed (experimentally or numerically) close to the superfluid-Bose glass quantum phase transition in three dimensions. Yao et al. { announce to have resolved} the { contradiction between the predictions of Fisher et al. [Phys. Rev. B {\bf 40}, 
546 (1989)] that $\phi = \nu z \geq 2$ (where $\nu$ and $z$ are critical exponents at the quantum phase transition) and recent experiments and simulations, showing an exponent $\phi \sim 1.1$.} Yao et al's resolution hinges on claiming that the $\phi \sim 1.1$ observations, which were conducted by varying the chemical potential, were not made to sufficiently close to the critical point to observe universal behavior. 
%{ On the other hand, a $\phi \sim 2.7$ (consistent with $\phi = \nu z$)} can be observed in their newly reported Quantum Monte Carlo simulations that vary disorder at fixed chemical potential. 
Here we critically examine their criteria for observing universal behavior. We show that past simulations were in fact conducted closer to the quantum critical point than the new results of Yao et al.; and that Yao et al.'s own results for varying chemical potential are consistent with previous results showing $\phi \sim 1.1$, and do not show a clear crossover to a different $\phi$ value. 
\end{abstract}
\maketitle

\section{Introduction}

 In a recent preprint \cite{Yaoetal2014}, Yao et al. reported large-scale quantum Monte Carlo (QMC) simulations of the superfluid-Bose glass (SF-BG) transition for two different models of disordered bosons in three dimensions: a) hardcore bosons with a random chemical potential uniformly distributed over a box $[-\Delta,\Delta]$, and whose transition is driven either by a varying average chemical potential $\mu$, or by disorder at fixed, \emph{zero} average chemical potential; b) a link-current model with the same form of disorder, and with fixed \emph{zero} average chemical potential.  They report a striking difference in the scaling of the critical temperature $T_c$ for condensation  as a function of the distance from the $T=0$ quantum critical point - QCP - $(\mu_c,\Delta_c)$, when the latter QCP is traversed following two different protocols

 \begin{itemize}
 \item protocol (1): at fixed $\Delta$ and variable chemical potential $\mu$; 
 \item protocol (2): at fixed, \emph{zero} $\mu$ and variable $\Delta$. 
 \end{itemize}
 In particular { they report that}, following the protocol (1), $T_c \sim |\mu-\mu_c|^\phi$, with $\phi \approx 1.1$, while, following the protocol (2), $T_c \sim |\Delta-\Delta_c|^{\phi}$, with $\phi\approx 2.7$. Yao et al. { attribute} this strong difference { to} the fact that protocol (2) is indeed observing the correct, asymptotic scaling of $T_c$ as $T\to 0$, because the whole protocol operates at \emph{fixed average density} $n=1/2$ due to the average particle-hole symmetry associated with the choice $\mu=0$. 
 On the other hand the scaling behavior observed with protocol (1) is interpreted as a ``transient" behavior { (in the words of Yao et al.)}, { by which they mean that the behavior is not universal because it is too far from the quantum critical point.  In particular the true scaling behavior would not be revealed because the density changes significantly with respect to the finite density $n_c$ at the QCP when the chemical potential is varied}. 
 Most importantly, the scaling theory of the SF-BG transition of Fisher et al. \cite{Fisheretal1989} postulates that the universal scaling function of the free energy around the QCP depends on the temperature only through the ratio $T/|\delta|^{\nu z}$ ($\delta$ being the distance to the QCP), implying that $\phi = \nu z$. This relationship is verified by the estimates of the critical exponents of Yao et al.  \cite{Yaoetal2014}, $\nu = 0.88(5)$ and $z=d=3$, consistent { (only roughly for $\nu$)} with previous estimates in the literature \cite{HitchcockS2006, Yuetal2012b}. The agreement with the simplest possible scaling theory of the SF-BG transition \cite{Fisheretal1989} is therefore a tantalizing aspect to conclude that the exponent observed with the protocol (2) is indeed the correct one. 
 
 Remarkably, an exponent $\phi\approx 1.1(2)$, consistent with what Yao et al. observe within protocol (1), has been reported in recent numerical studies on quantum spin realizations of the SF-BG transition \cite{Yuetal2010, Yuetal2012b, Yuetal2012a} with significantly different forms of disorder (site dilution vs. bond and anisotropy disorder), as well as in a series of recent experiments on doped quantum magnets in a magnetic field (acting as chemical potential), supposed to possess the same symmetries as those of a lattice-boson Hamiltonian \cite{Yuetal2012a, Yamadaetal2011, ZheludevH2011, Huvonenetal2012}. All the above cited simulations and experiments have been performed using protocol (1). Hence the conclusions of Yao et al. \cite{Yaoetal2014} would imply that the above studies are affected by significant limitations, and are only observing effective exponents, as they do not approach the QCP sufficiently close to avoid the effects of a varying density. 
  
  \medskip
   In the present comment we would like to point out the following { issues}:
   
\begin{enumerate}  

 \item Yao et al. do not provide convincing evidence that the $\phi\approx 1.1$ exponent obtained by them via the protocol (1) is a ``transient" one. In fact { a more detailed analysis shows that their data consistently indicate an exponent $\phi = 1.2(1)$}. Overall their simulations agree with the previous studies performed with protocol (1), {which strengthens the argument} that an exponent $\phi\approx 1.1$ might be a robust feature of the onset of $T_c$ around the SF-BG QCP when tuning the transition via the chemical potential;    
 
 \item the criterion of a (nearly) fixed density across the transition (hereafter called the \emph{density criterion}) is critically re-examined. We show that the data of Ref.~\onlinecite{Yuetal2012a} are in fact complying with the criterion $\epsilon_n = |n/n_c-1|\ll 1$ over a temperature range sufficiently broad to rule out an exponent $\phi \sim 2.7$; { moreover, they are fully} complying with the conventional criterion for the proximity to the QCP, imposed on the actual driving parameter of the transition, $\epsilon_{\mu} = |\mu/\mu_c-1| \ll 1$. In the case of the disorder-driven transition of protocol (2), one should equivalently apply a "disorder criterion" $\epsilon_\Delta =  |\Delta/\Delta_c-1| \ll 1 $. In fact, using the smallness of the parameters $\epsilon_n$, $\epsilon_{\mu}$ and $\epsilon_{\Delta}$ as a quality factor for the numerical simulations of the scaling of $T_c$ around the QCP, we find that the results of  Ref.~\onlinecite{Yuetal2012a} are of comparable, better, or even substantially better quality than the results of Yao et al.
 
 %\item we critically re-examine the suggestion of Yao et al. that \emph{fixing} the density upon crossing the QCP is a secure strategy to detect the correct $\phi$ exponent. Indeed we show that a fixed-density trajectory along the critical surface $T_c(\mu,\Delta)$ might exhibit a value of $\phi$ which is inconsistent with the observation of Yao et al. and Fisher's scaling theory \cite{Fisheretal1989} for protocols changing the chemical potential, namely for any protocol \emph{except} protocol (2).  

 \end{enumerate}

The aim of this comment is not to criticize the { Quantum Monte Carlo simulations} of Yao et al., which are undoubtedly solid, but rather their conclusions. 
 %that Yao et al. draw from their results. Nonetheless we find that the results of Yao et al., along with the results we present here, expose the considerable complexity of the problem of disordered interacting bosons, etc. etc. 

\begin{figure}[h!]
\begin{center}
\includegraphics[width=8cm]{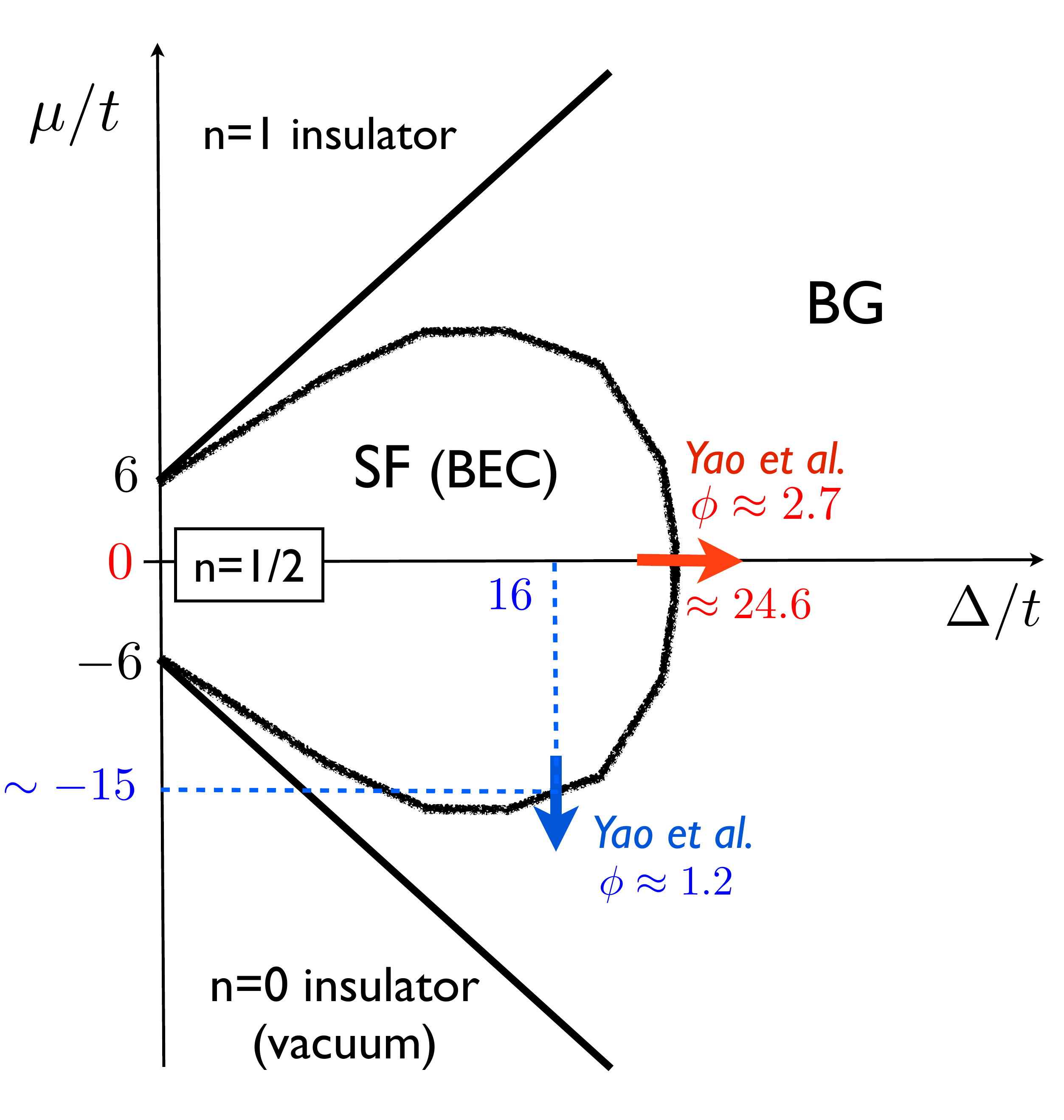}
\caption{Schematic $T=0$ phase diagram of 3D hardcore-boson in a random chemical potential, with the location of the SF-BG transitions (blue and red arrow, following protocols (1) and (2) respectively) considered by Yao et al.~.}
\label{f.hcb}
\end{center}
\end{figure}

\section{Putting existing results into context}

\subsection{Hardcore bosons}

 Our first remark concerns the global phase diagram of the model of disordered bosons investigated by Yao et al., and the specific locations of the SF-BG transitions that Yao et al. studied in details. 
 The Hamiltonian of the model is 
 \begin{equation}
 {\cal H} = -t \sum_{\langle ij \rangle} \left( b_i^{\dagger} b_j + {\rm h. c.} \right) - \sum \mu_i n_i
 \label{e.hc}
 \end{equation}
 where $b$, $b^{\dagger}$ are hardcore-boson operators, $\mu = \mu + \delta\mu_i$ and $\delta\mu_i$ is randomly distributed within the box $[-\Delta,\Delta]$.
 The $T=0$ phase diagram of the above model - as emerging from well-known results and from the results of Yao et al. - is schematized in Fig.~\ref{f.hcb}; notice the similarity to the phase diagram of free fermions in a random box potential \cite{Bulkaetal1987,Markos2006}. We observe that while protocol (1) - as described in the introduction - corresponds to crossing the SF-BG transition at a rather ``generic" transition point, protocol (2) corresponds to a  { very special point} in the phase diagram, namely it lies along the line $\mu=0$, at which the average density $n$ is fixed at $n=1/2$ for \emph{any value of the disorder} and \emph{for any finite temperature}. This property is related to  an \emph{average} particle-hole symmetry - not an exact symmetry on any finite-size system, but a symmetry valid on average over disorder, and, due to the self-averaging nature of the density, an \emph{exact} symmetry in the thermodynamic limit. As we will further point out, the existence of such a special point is { closely} related to the hardcore nature of the bosonic particles. Due to its special location at the tip of the SF lobe in the phase diagram, the transition point represents the SF-BG transition taking place at \emph{maximum strength of disorder} for the model in question, and taking place for the \emph{maximum value of the density} (here to be understood as density \emph{modulo} one,  whence $n=1/2$ is the maximum possible value.)  The observation by Yao et al. of an exponent $\phi \approx 2.7$ consistent with the relationship $\phi = \nu z$ is limited to this particular point in the phase diagram.

\begin{figure}[h!]
\begin{center}
\includegraphics[width=10cm]{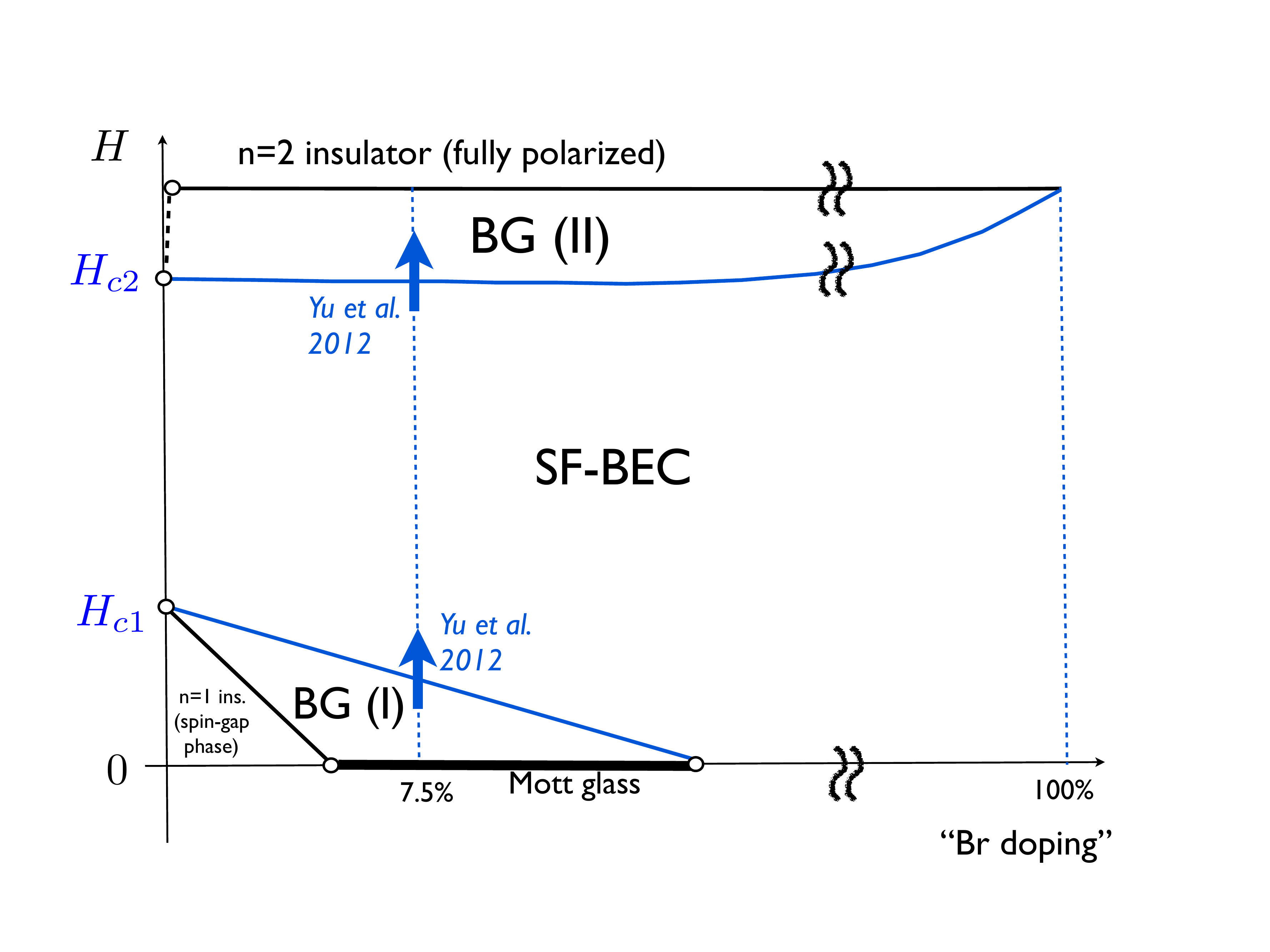}
\caption{Schematic phase diagram of the \emph{theoretical model} of Br-doped DTN, with the positions of the SF-BG transitions (blue arrows) studied in Refs.~\onlinecite{Yuetal2012a,Yuetal2012b}~.}
\label{f.DTN}
\end{center}
\end{figure}

\subsection{Model for Br-doped DTN}

Refs.~\onlinecite{Yuetal2012a,Yuetal2012b} have considered on the other hand a theoretical model of disordered $S=1$ spins, capturing most of the features of the magnetic behavior of doped dichloro-tetrakis-thiourea Nickel (DTN). In particular, Br doping of DTN is modeled via a correlated bimodal distribution of the magnetic bonds along the crystallographic $c$-axis and of the single-ion anisotropies \cite{Yuetal2012a}. The resulting Hamiltonian reads
\begin{eqnarray}
{\cal H}_{\rm Br-DTN} &=& \sum_{\langle ij \rangle_{c}}  
J_{c,\langle ij \rangle} ~{\bm S}_{i}\cdot{\bm S}_{j} ~~+ 
J_{ab} \sum_{\langle lm \rangle_{ab}} 
{\bm S}_{l}\cdot{\bm S}_{m} \nonumber \\
&+& \sum_{i} D_i (S^z_{i})^2
- g\mu_B H \sum_{i}  S^z_{i}.
\label{e.Ham}
\end{eqnarray}
Here $S_i^{\alpha}$ ($\alpha=x,y,z$) are $S=1$ spin operators, coupled on a cubic lattice with antiferromagnetic  interactions $J_c$ along the $c$ axis and $J_{ab}$ in the $ab$ plane; $D$ is a strong single-ion anisotropy. In pure DTN $J_{c}=2.2$ K, $J_{ab}=0.18$ K, $D=8.9$ K.\cite{Zvyaginetal2007} Doping at a concentration $p$ is modeled as increasing the $J_c$ bond strength to $J_c'=2.3 J_c$ on a fraction $2p$ of the $c$-axis bonds, and reducing the single anisotropy $D$ to $D'=D/2$ on one of the two ions connected by the bond. In particular we find that $g=2.31$ allows { this model to reproduce} the magnetization curve of Br-DTN at $p\approx 8\%$ \cite{Yuetal2012a}. When mapped \cite{Yuetal2012a} onto softcore bosons with maximum occupancy $n=2$, the above model provides a valuable example of a 3D dirty-boson system, for which a SF-BG transition can be controlled e.g. by varying the { magnetic} field (which acts as a chemical potential) or by varying the disorder. { In experiments, only the magnetic field can be continuously tuned, however in the theoretical model we can continuously tune the { Br substitution} from 0 to 100\%, leading to the phase diagram that we { sketch} in Fig.~\ref{f.DTN}.} There the SF-BG transitions explored in Ref.~\onlinecite{Yuetal2012a, Yuetal2012b} are two protocol-(1) transitions tuned by the magnetic field /chemical potential.
As we will discuss later in more details, the two transitions occur at very different densities: a very low density $n \sim 10^{-4}$ around the lower critical field $H_{c1}$, and a more sizable density $n \sim 0.1$ at the upper critical field $H_{c2}$.  
Due to the special nature of the disorder induced by ``theoretical" Br doping - that of lowering the local spin-gap around the dopants -  one finds that doping enhances the  
SF region of the phase diagram. Therefore a transition point of maximum density ($n\sim 1/2$) and maximum disorder strength as in the hardcore boson system is not present in the model for Br-DTN. An \emph{exact} particle-hole symmetry is featured along the $H=0$ line, where $n=1$ (on \emph{each} site of the lattice for any size, temperature or disorder strength). Along such line the BG is replaced by a Mott-glass phase, whose transition to a SF belongs potentially to a different universality class than the SF-BG transition, and therefore we shall not consider it here.

\begin{figure}[h!]
\begin{center}
\mbox{
\includegraphics[width=4cm]{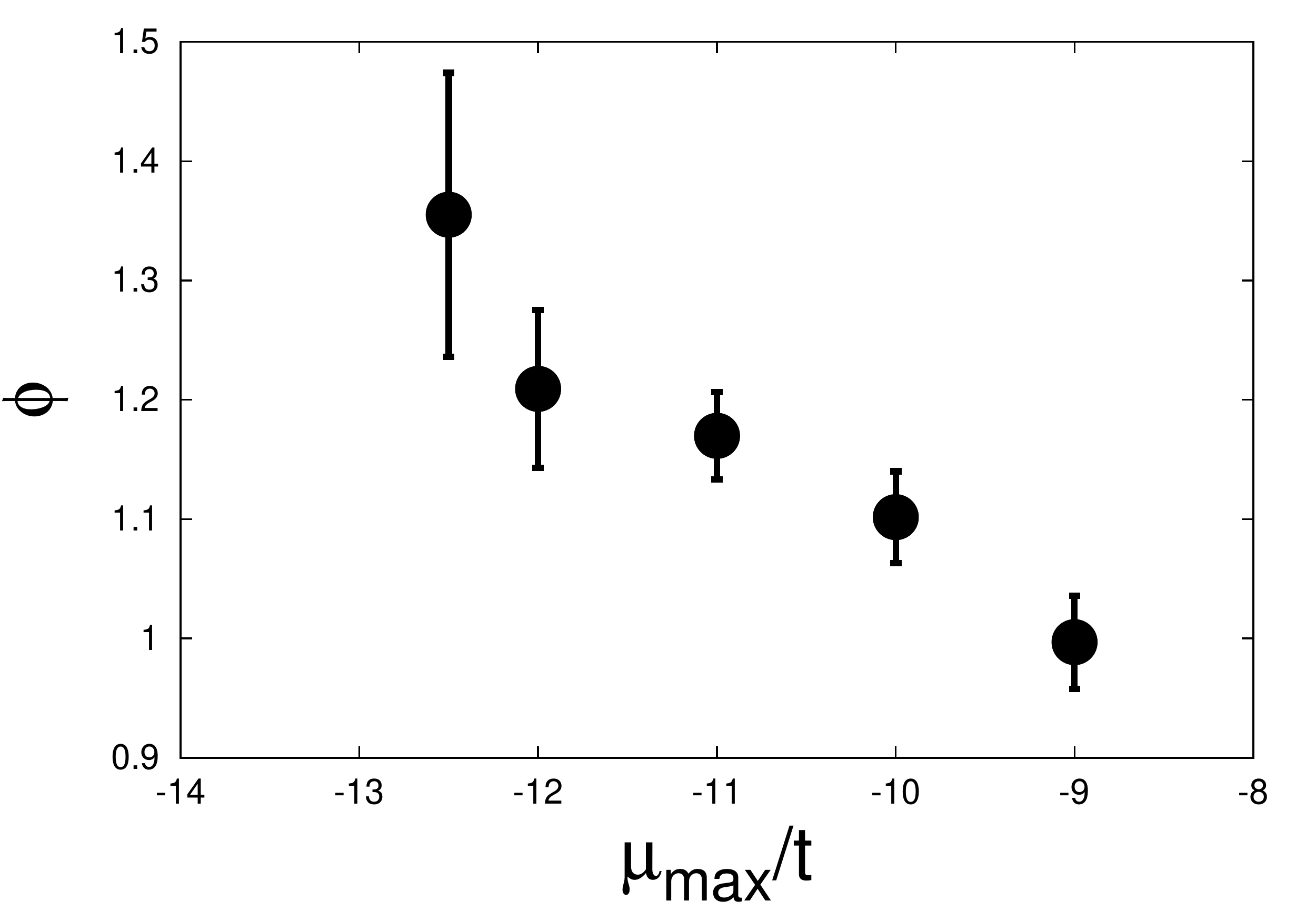}
\includegraphics[width=4cm]{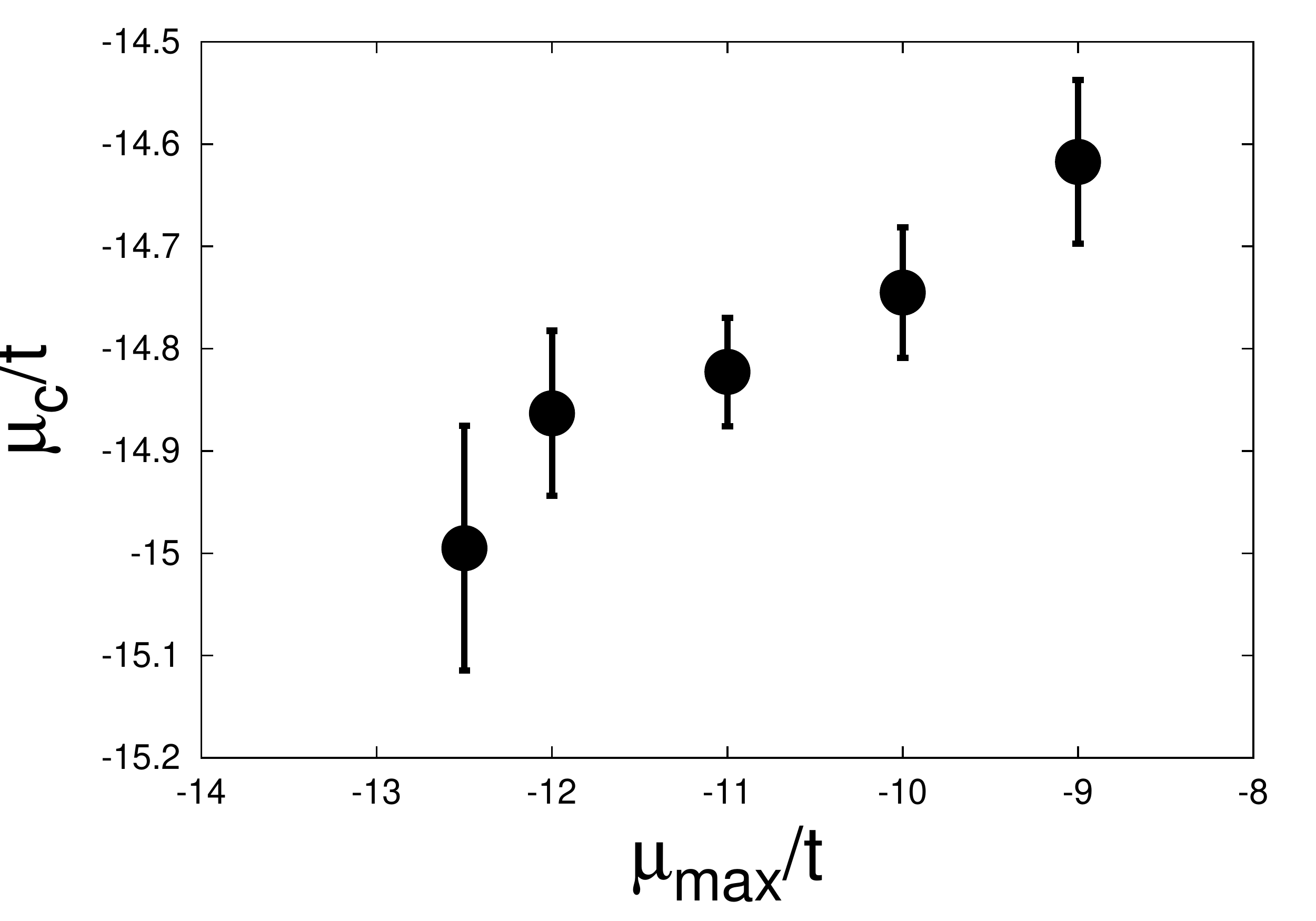}}
\includegraphics[width=6cm]{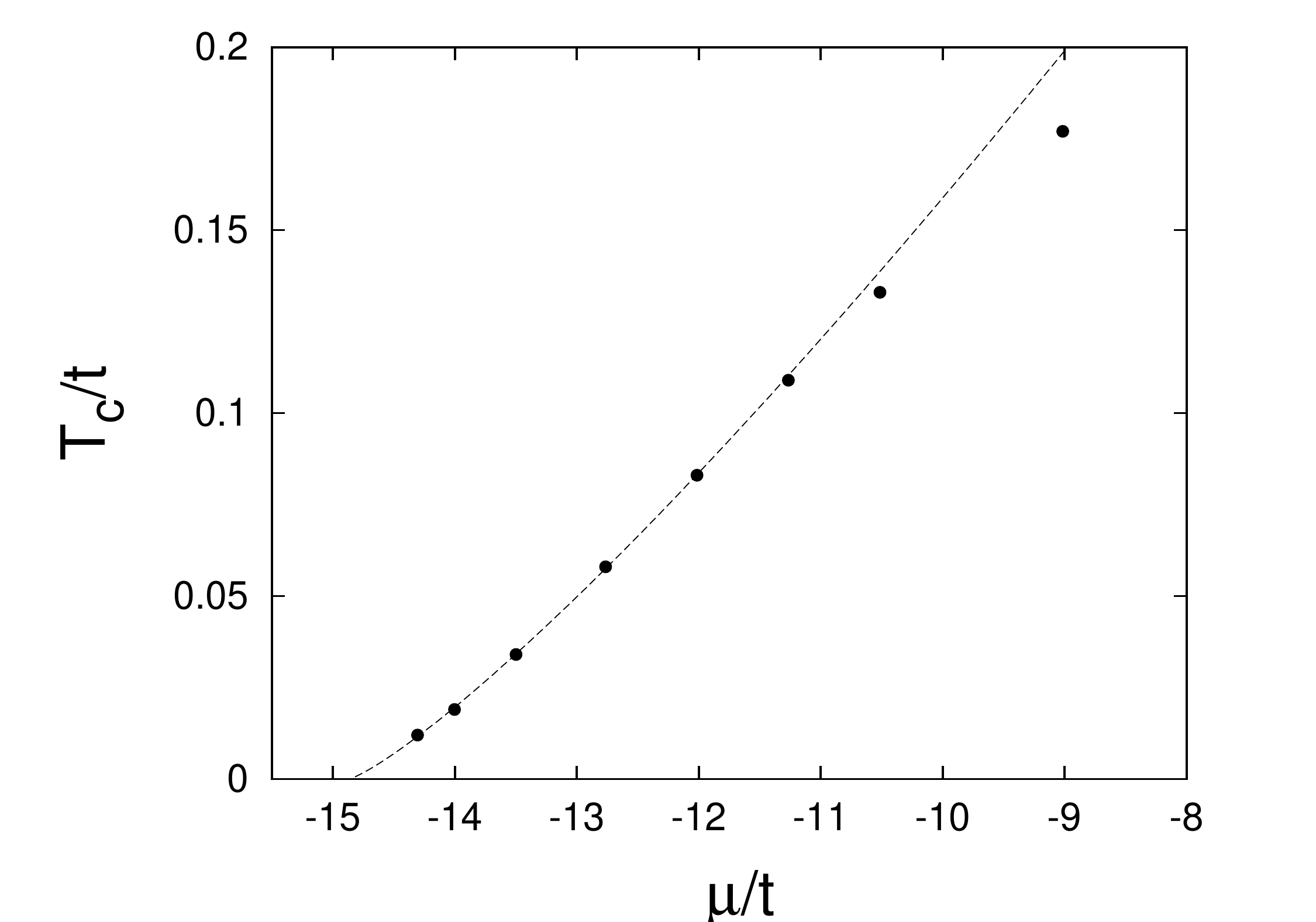}
\caption{Windowing analysis of the $T_c$ vs. $\mu$ data of the protocol-(1) transition of Yao et al. The last plot shows the fit with $\mu_{\rm max}/t = -12$.}
\label{f.window}
\end{center}
\end{figure}

\section{Windowing analysis of Yao et al.'s results}

Yao et al. fit their $T_c$-vs.-$\mu$ data at the protocol-(1) transition for the Hamiltonian Eq.~\eqref{e.hc} with the form $A*(\mu-\mu_c)^{\phi}$ with $A$ and $\mu_c$ as fitting parameters, and $\phi$ \emph{fixed} at the value 1.1. They conclude that their data are inconsistent with this fitting form because the lowest-temperature point falls { slightly} outside the fitting curve. { Their mistake is in assuming that $\phi$ is precisely 1.1. In previously simulations and experiments, $\phi \sim 1.1$ with different error bars around 10\%. With a slightly different choice of $\phi$ within that window of error bars, their data fits to a single power law with no deviation (Fig.~\ref{f.window}). Furthermore, we perform a windowing analysis on their result to extrapolate the exponent to the critical point. }
A fitting analysis over a variable window of chemical potentials $[\mu_{\rm min}, \mu_{\rm max}]$ (where $\mu_{\rm min}$ is set to the minimum chemical potential of the data, and $\mu_{\rm max}$ is varied) shows that their results are indeed consistent with a $\phi\approx 1.2(1)$, in agreement with the previous analyses \cite{Yuetal2012a,Yuetal2010}  of SF-BG transitions following protocol (1) performed on different models - see Fig.~\ref{f.window}.

{ Thus taking Yao et al's data together with previous results,} a $\phi$ exponent appears to emerge for widely different models with different forms of disorder: a random-box chemical potential in Ref.~\onlinecite{Yaoetal2014}, bond and single-ion-anisotropy randomness in Ref.~\onlinecite{Yuetal2012a}, site dilution in Ref.~\onlinecite{Yuetal2010}. This would suggest that such an exponent is { a robust} feature of the transition. 

The conclusion of Yao et al. is rather different: they interpret their protocol-(1) data as showing a ``transient" exponent, due to the fact that the density of bosons changes significantly along the $T_c$-vs-$\mu$ critical line. { They conclude that for transitions with changing density, the correct $\phi$ exponent is revealed only} for temperatures $T$ and chemical potentials $\mu$ such that $\epsilon_n = |n(\mu,T)/n_c-1| \ll 1$, where $n_c = n(\mu_c,T=0)$ is the density at the quantum critical point. In particular, Yao et al. concude that, given that \emph{their} data with variable density show an exponent $\phi\sim 1.1$ and do not comply with the criterion $\epsilon_n\ll 1$, then \emph{all} previous experiments and calculations consistent with the same exponent must be affected by the limitation of having a variable density. In the following { we challenge the { criterion} $|n(\mu,T)/n_c-1| \ll 1$, and show additional unpublished data even closer to the quantum critical point that supports the exponent $\phi \sim 1.1$.}

\begin{figure}[h!]
\begin{center}
\includegraphics[width=7cm]{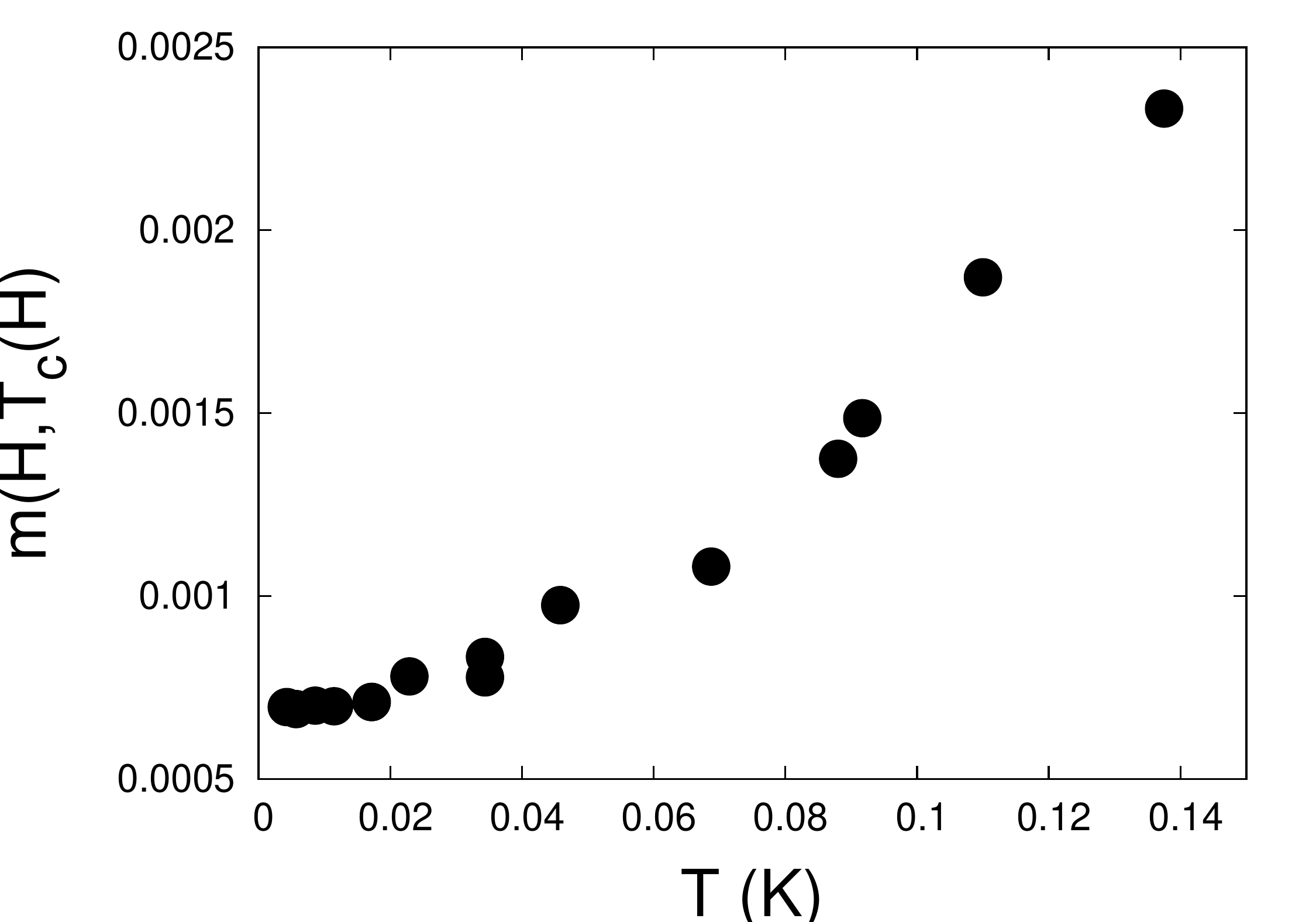}
\includegraphics[width=7cm]{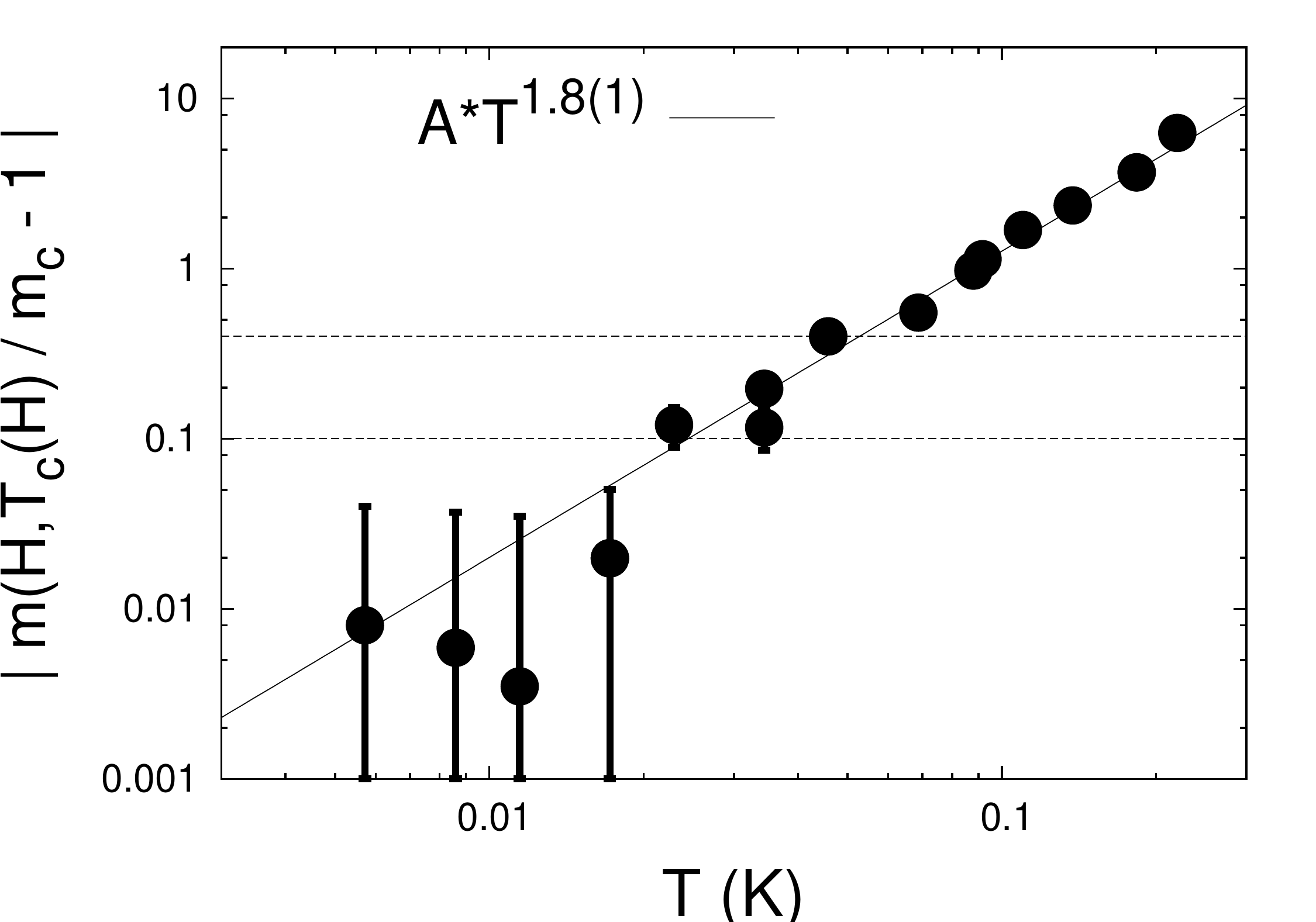}
\caption{Density criterion analysis of the data of Ref.~\onlinecite{Yuetal2012a} around $H_{c1}$. In the lower panel the horizontal dashed lines mark the 40\% threshold and the 10\% threshold (upper and lower line respectively).}
\label{f.dcHc1}
\end{center}
\end{figure}

\begin{figure}[h!]
\begin{center}
\includegraphics[width=7cm]{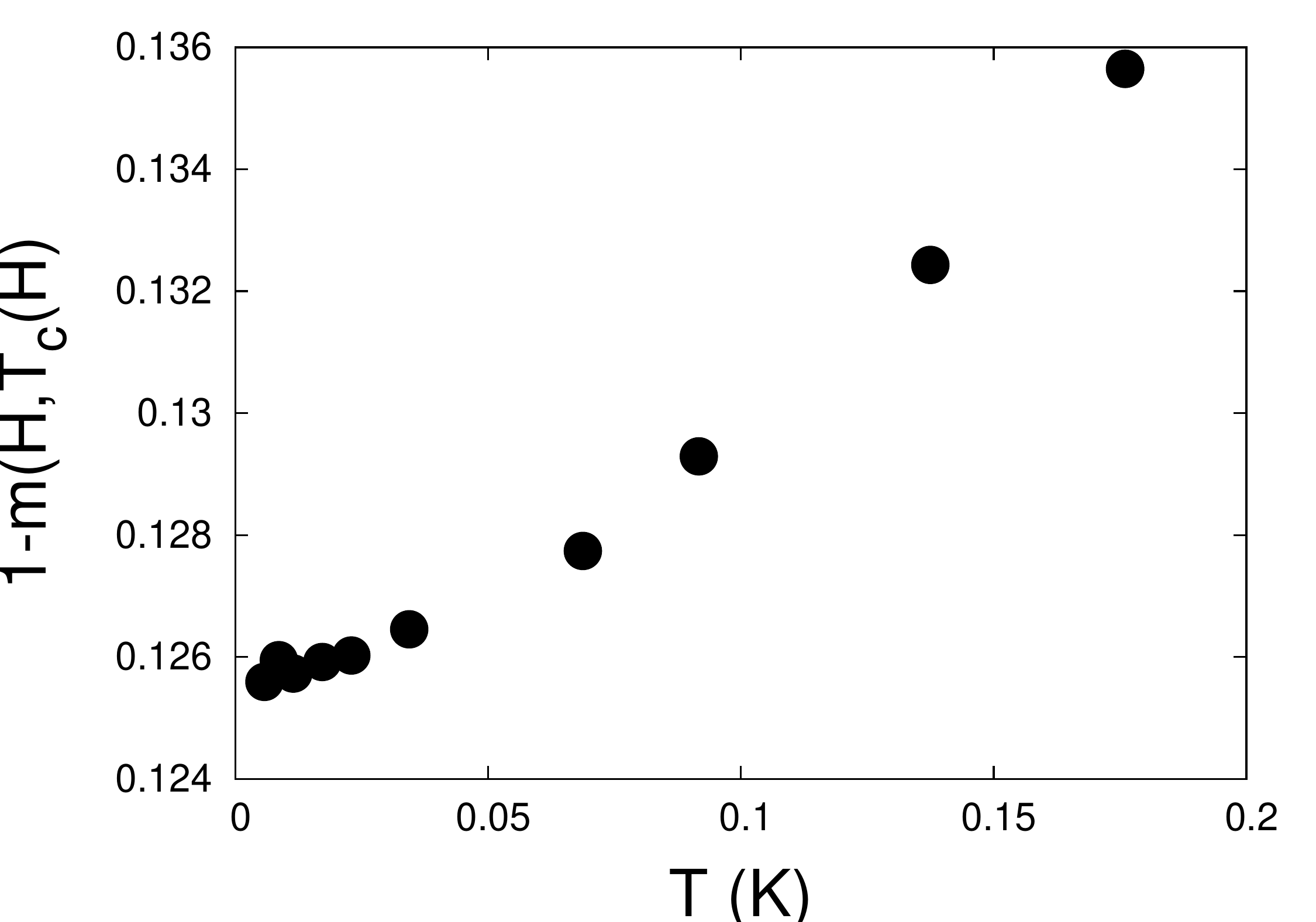}
\includegraphics[width=7cm]{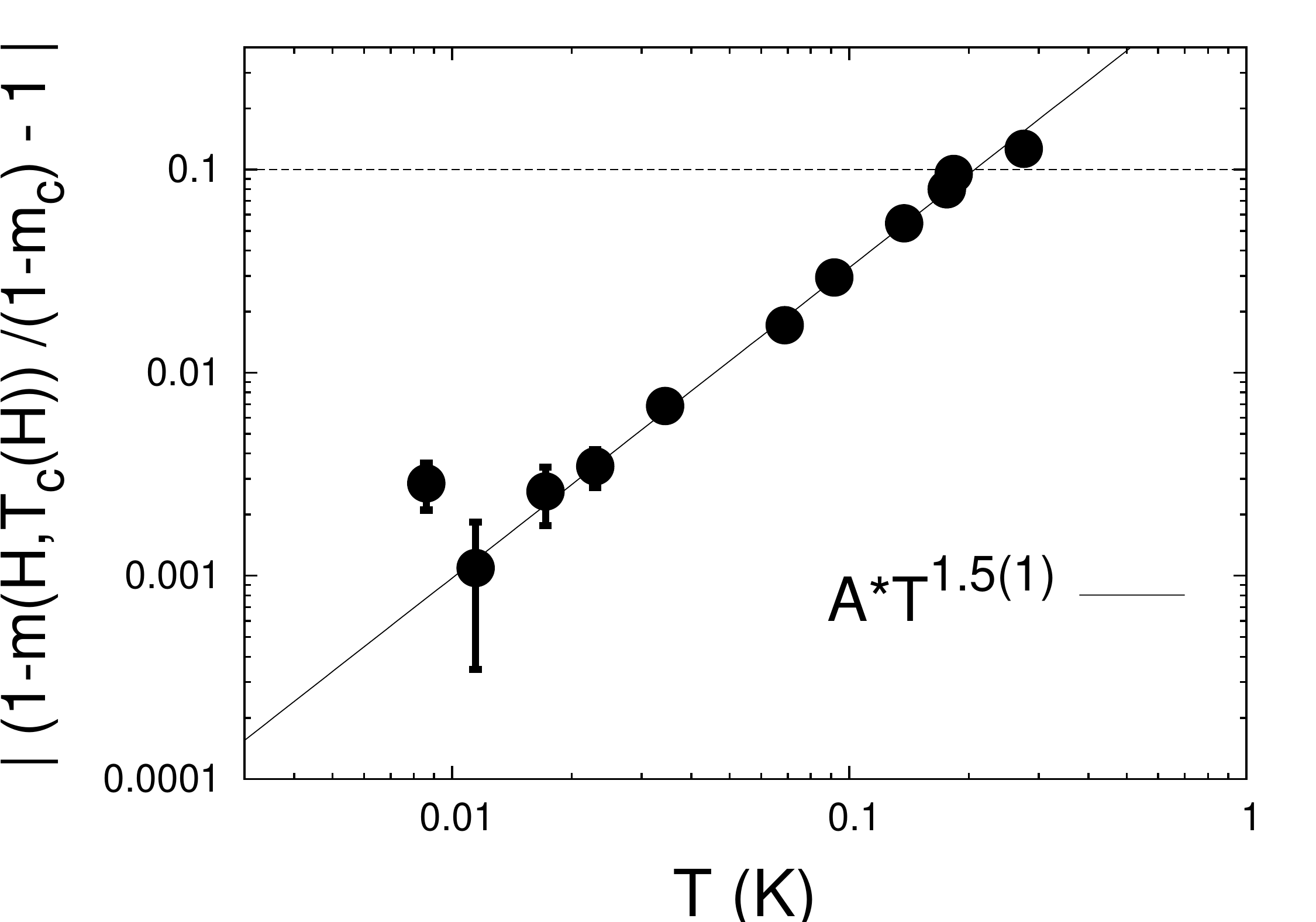}
\caption{Density criterion analysis of the data of Ref.~\onlinecite{Yuetal2012a} around $H_{c2}$. In the lower panel the horizontal dashed line marks the 10\% threshold.}
\label{f.dcHc2}
\end{center}
\end{figure}

\section{Density criterion applied to the theory results for Br-DTN}

A common criterion to define proximity to the QCP crossed at fixed disorder and variable chemical potential (protocol (1)) would be $|\mu/\mu_c-1| \ll 1$, which is indeed satisfied by the protocol-(1) data of Yao et al., as well as by the data of Refs. \onlinecite{Yuetal2010,Yuetal2012a,Yuetal2012b}. The density criterion of Yao et al., on the other hand, imposes that not only the driving parameter ($\mu$) be close to the critical value $\mu_c$, but that also its conjugate variable ($n$) be close to its value $n_c$ at criticality. This is a rather unusual criterion when applied to known, more standard transitions than the one at hand. It turns out to be much more restrictive than the conventional criterion - for instance, such a criterion would imply that the critical region is of \emph{zero width} for transitions with a divergent compressibility, such that an infinitesimal change of chemical potential away from the critical point entails a substantial change in the density. 

 Nonetheless, the density criterion of Yao et al. would justify the difference in the $\phi$-exponent estimates between transitions following protocol (1) vs. protocol (2). In the latter protocol, as already explained above, the average density is strictly constant for any disorder strength and temperature. Here we re-examine the QMC data of Refs.~\onlinecite{Yuetal2012a} for the model Eq.~\eqref{e.Ham} in the light of the density criterion, and we show that \emph{data satisfying the density criterion do not in turn exhibit a $\phi$ exponent consistent with $\phi = \nu z$.} In particular the data of Ref.~\onlinecite{Yuetal2012a} lie closer to the QCP (according to the density criterion) than the data of Yao et al. following protocol (2) (according to an equivalent disorder criterion). Hereafter we cite the temperature and magnetic field data for the theory model Eq.~\eqref{e.Ham} in \emph{physical units}, in order to connect to the experimental data of Br-DTN as well - even though we shall only exploit theory data. 
 
 \subsection{Magnetization along the $T_c(H)$ critical line}
 
  In the case of the spin model for Br-DTN, Eq.~\eqref{e.Ham} the density criterion takes the form of a \emph{magnetization} criterion, $\epsilon_m = |m/m_c-1| \ll 1$, where $m=\langle S_i^z \rangle$. Moreover, close to the saturation field, the density criterion must be { described} in terms of the distance to the saturation magnetization $1-m$, namely $\epsilon_m = |(1-m)/(1-m_c)-1| \ll 1$. 
Figs.~\ref{f.dcHc1} and \ref{f.dcHc2} show the magnetization and the $\epsilon_m$ parameter along the critical line $T_c(H)$  at finite temperature for the model of Eq.~\eqref{e.Ham} both in the vicinity of the lower critical field $H_{c1} = 1.172(5)$ T, and of the upper critical field $H_{c2} = 12.302(5)$ T (see Ref.~\onlinecite{Yuetal2012a}). Here $m_c$ is estimated as the density at the lowest accessed temperature, and for a system size $18^3$ - we check that the data in question are essentially devoid of thermal as well as finite-size effects. A first observation is that the two QCPs correspond to two very different density/magnetization regimes: while the density at $H_{c1}$ is very weak, $m_c \approx 7*10^{-4}$, the density around $H_{c2}$ is significantly bigger, $m_c \approx  0.126$. In both cases the magnetization is found to follow approximately a power law along the critical curve, $m(H,T_c(H)) \approx m_c + A*T_c(H)^{x_m}$ { with} $x_m \approx 1.8(1)$ close to $H_{c1}$ and   $x_m \approx 1.5(1)$ close to $H_{c2}$ (a similar power law has been observed along the quantum critical trajectory above $H_{c1}$ in Ref.~\onlinecite{Yuetal2012b}). 
We observe that the data for the critical line reported in Ref.~\onlinecite{Yuetal2012a} can indeed access regions with $\epsilon_m \ll 1$, where the correct scaling of $T_c$ vs. $H$ is supposed to be observed according to the density criterion of Yao et al. Indeed around $H_{c1}$ our data clearly cover the region $\epsilon_m < 10\%$ (corresponding to $T \lesssim 23$ mK)  while around $H_{c2}$ they { even} access the region $\epsilon_m < 1\%$ (corresponding to $T < 50$ mK). Remarkably, the experimental data of Ref.~\onlinecite{Yuetal2012a} have access as well the above cited regions. 

\subsection{$T_c(H)$ scaling close to the critical fields}

\begin{figure}[h!]
\begin{center}
\includegraphics[width=9cm]{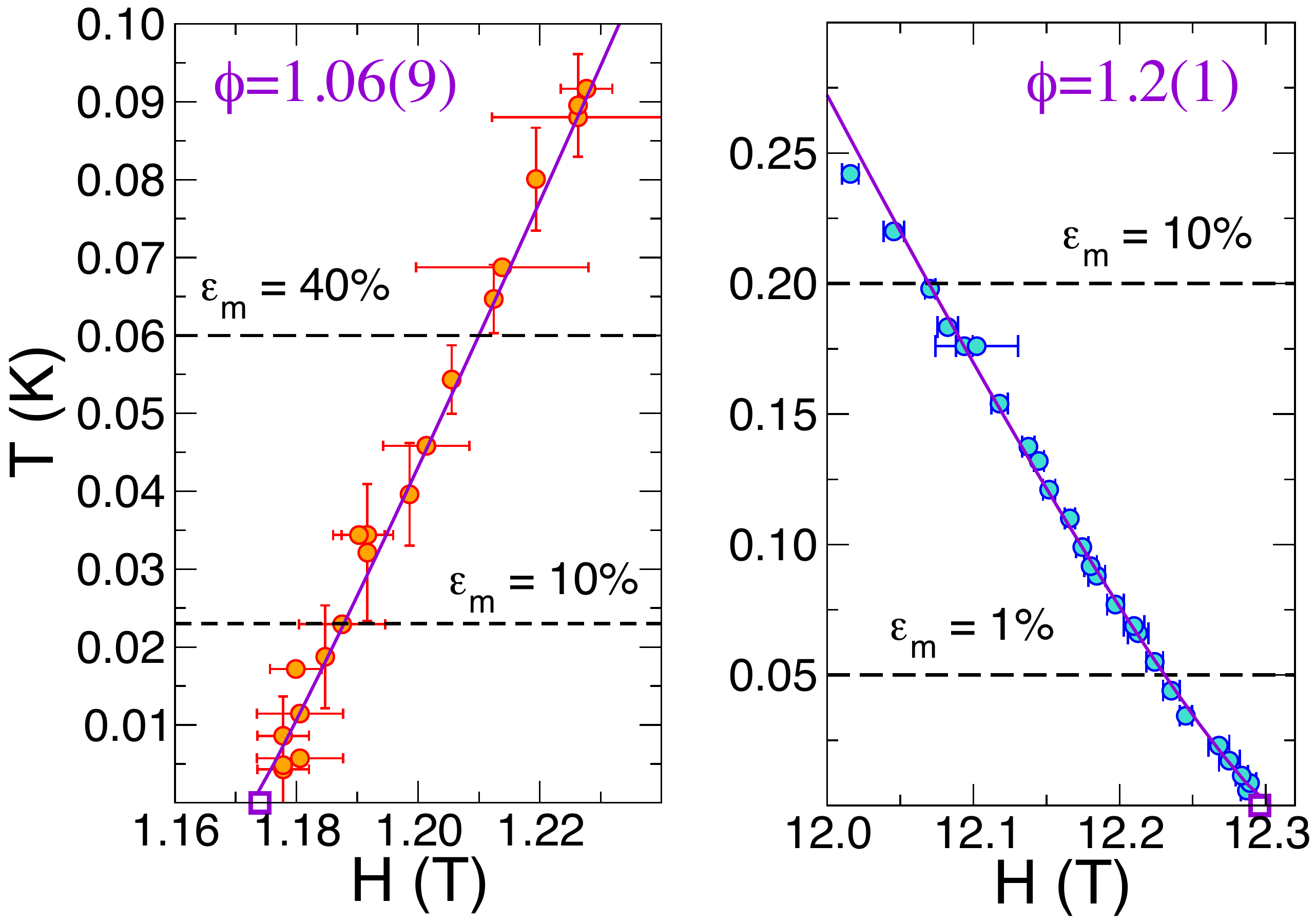}
\caption{Critical curves $T_c$ vs. $H$ around $H_{c1}$ (left panel) and $H_{c2}$ (right panel) for the model of Br-doped DTN. The dashed lines indicate relevant thresholds for the density (magnetization) criterion. The solid lines are fits to $T_c = A*|H-H_c|^{\phi}$, with $\phi$ exponents in the legend. The open squares represent the estimates of the QCP $H_ {c1}$ and $H_{c2}$ from $\beta$-doubling QMC data (see the text).}
\label{f.dcTc}
\end{center}
\end{figure}

Fig.~\ref{f.dcTc} shows the critical lines around $H_{c1}$ and $H_{c2}$ for the model Eq.~\eqref{e.Ham} with the threshold values for the density-criterion parameter $\epsilon_m$. We observe that, despite their sizable error bars, data close to $H_{c1}$ cannot be reconciled with an exponent $\phi= \nu z \approx 2.7$ even in the region in which $\epsilon_m < 10\%$. The latter conclusion applies even more strongly to the data close to $H_{c2}$ for an $\epsilon_m$ as small as $1\%$. 
In particular in Ref.~\onlinecite{Yuetal2012a} a $\phi$ exponent of $1.2(1)$ close to $H_{c2}$ was extracted from data with $\epsilon_m \lesssim 5\%$.  There is no clear sign of a change of scaling upon reducing the temperature. This suggests that one observes a $\phi$ exponent { that} is consistently different from the prediction $\phi = \nu z$ even when complying with the density criterion. 
  
   The observation of an exponent $\phi=1.1-1.2$ is obtained for SF-BG transitions occurring at critical densities (magnetizations) $n_c$ ($m_c$) which differ by almost three orders of magnitude between $H_{c1}$ and $H_{c2}$ - a remarkable fact in our opinion, which also shows how Br-DTN can give access to two SF-BG transitions in rather different regimes. Finally the reliability of the $\phi$-exponent estimates reported in Ref.~\onlinecite{Yuetal2012a} is further supported by the fact that the position of the QCPs at $H_{c1}$ and $H_{c2}$, obtained by extrapolation of the finite-temperature data of Fig.~\ref{f.dcTc}, is \emph{completely consistent} with an independent estimate that is provided by a scaling analysis of QMC data \cite{Yuetal2012b, Yuetalunp} obtained via an exponential cooling protocol ($\beta$-doubling scheme \cite{Sandvik2002}). The latter approach enables { us} to systematically eliminate  thermal effects from finite-size QMC data, without any \emph{a priori} assumption on the dynamical critical exponent $z$ at the QCP.

Concerning the distance to the critical field $H_{c1}$ and $H_{c2}$, $\epsilon_H = |H/H_c-1|$, the data used to extract the $\phi$ exponent in Ref.~\onlinecite{Yuetal2012a} had $\epsilon_H  \lesssim 5\%$ close to $H_{c1}$ and $\epsilon_H  \lesssim 2\%$ close to $H_{c2}$.
 In this respect, the data of Yao et al. for protocol (2) are of lower quality according to the disorder parameter for the distance to the critical point, $\epsilon_{\Delta} = | \Delta/\Delta_c-1|$. In fact the fit which leads to the estimated exponent $\phi = 2.7$ is based on 4 points in a region with $\epsilon_\Delta \lesssim 30 \%$ for the link-current model (Fig. 5 of Ref.~\onlinecite{Yaoetal2014})  and on 4 points in a region which has again $\epsilon_\Delta \lesssim 30 \%$ for the hardcore-boson model (Fig. 6 of Ref.~\onlinecite{Yaoetal2014}). Hence we can conclude that the existing numerical data supporting $\phi=1.1-1.2$ are of \emph{better} quality - according to the $\epsilon_{\mu}$ parameter as well as to the (more restrictive!) density criterion based on the $\epsilon_{m(n)}$ parameter - than the data produced by Yao et al. in support of $\phi = 2.7$ - according to the $\epsilon_{\Delta}$ parameter.

\section{Conclusions}

In this note we have shown that previous theoretical data for a model of Br-doped DTN - supporting a scaling of the critical temperature close to a 3d SF-BG transition with an exponent $\phi\approx 1.1-1.2$ - satisfy the { criterion} of a nearly constant density which Yao et al. in Ref.~\onlinecite{Yaoetal2014} set as conditions for the observation of the actual asymptotic $\phi$ exponent. Therefore the conclusion of the latter reference -- that the observed value $\phi\approx 1.1-1.2$ be a ``transient" value on the way to a value consistent with the relation $\phi = \nu z$ --  is actually invalid on the basis of the data at hand. While our analysis strictly applies to the theory data of Ref.~\onlinecite{Yuetal2012a}, we can conjecture that a similar conclusion is also valid for much of the experimental data supporting a similar value of $\phi$ (typically $\phi \sim 1.1$) for other magnetic compounds which are strong candidates for the realization of the 3d SF-BG transitions \cite{Yamadaetal2011,ZheludevH2011,Huvonenetal2012}.  

To convincingly conclude that  $\phi \sim 1.1$ is a transient exponent, it would be necessary to { demonstrate} the crossover between the supposedly transient and the actual $\phi$ exponent.  The observation of Yao et al. of an exponent $\phi$ consistent  with the relationship $\phi = \nu z$ is quite remarkable, but it remains limited at present to the case of a special point in the phase diagram of hardcore bosons. In the light of the above discussion, Fisher et al.'s scaling Ansatz, predicting $\phi = \nu z$, remains insufficient to explain the body of existing numerical and experimental results. As elaborated in Ref.~\onlinecite{Yuetal2012b}, the violation of the prediction $\phi = \nu z$ implies unconventional scaling \emph{at finite temperature} only, whose origin still remains to be clarified theoretically.  Further work - both theoretical and experimental - is therefore necessary to reconcile the observations of Yao et al. at fixed density with the existing data at variable density. All these results further exhibit the complexity and the richness of the critical properties at the 3d SF-BG transition.    

\section{Acknowledgements}
We thank Z. Yao and N. Prokof'ev for useful discussions and correspondence.


\begin{thebibliography}{99}
\bibitem{Yaoetal2014} Z. Yao, K. P. C. da Costa, M. Kiselev, and N. Prokof'ev, arXiv:1402.5417 (2014).
 \bibitem{Fisheretal1989} M. P. A. Fisher, P. B. Weichman, G. Grinstein, and D. S. Fisher, Phys. Rev. B {\bf 40}, 
546 (1989). 
\bibitem{Yuetal2012b} R. Yu, C. F. Miclea, F. Weickert, R. Movshovich, A. Paduan-Filho, V. S. Zapf, and T. Roscilde, Phys. Rev. B {\bf 86}, 134421 (2012).   
\bibitem{HitchcockS2006} P. Hitchcock and E. S. S\o rensen, Phys. Rev. B {\bf 73}, 174523 (2006).

\bibitem{Yuetal2012a} R. Yu, L. Yin, N. S. Sullivan, J. S. Xia, C. Huan, A. Paduan-Filho, N. F. Oliveira Jr., S. Haas, A. Steppke, C. F. Miclea, 
F. Weickert, R. Movshovich, E.-D. Mun, B. S. Scott, V. S. Zapf, and T. Roscilde, Nature {\bf 489}, 379 (2012).  

\bibitem{Yuetal2010} R. Yu, S. Haas, and T. Roscilde, Europhys. Lett. {\bf 89}, 10009 (2010).

\bibitem{Yamadaetal2011} F. Yamada, H. Tanaka, T. Ono, and H. Nojiri,
Phys. Rev. B {\bf 83}, 020409 (2011). 
\bibitem{ZheludevH2011} A. Zheludev and D. H\"uvonen, Phys. Rev. B {\bf 83}, 216401 (2011).
\bibitem{Huvonenetal2012} D. H\"uvonen, S. Zhao, M. M\aa nsson, T. Yankova, E. Ressouche, C. Niedermayer, M. Laver, S. N. Gvasaliya, and A. Zheludev, 
Phys. Rev. B {\bf 85}, 100410(R) (2012).



\bibitem{Bulkaetal1987} B. Bulka, B. Kramer, and A. MacKinnon, Z. Phys. B: Condens. Matt. {\bf 60}, 13 (1985).
\bibitem{Markos2006} P. Marko\v s, Acta Phys. Slov. {\bf 56}, 561 (2006). 

\bibitem{Zvyaginetal2007}
S. A. Zvyagin, J. Wosnitza, C. D. Batista, M. Tsukamoto, N. Kawashima, J. Krzystek, V. S. Zapf, M. Jaime, N. F. Oliveira, Jr., and A. Paduan-Filho, 
Phys. Rev. Lett. {\bf 98}, 047205 (2007).
\bibitem{Yuetalunp} R. Yu et al., unpublished. 
\bibitem{Sandvik2002} A. W. Sandvik, Phys. Rev. B {\bf 66}, 024418 (2002).

\end{thebibliography}
\end{document}